\documentclass[final,3p,times]{elsarticle}

\usepackage{amssymb}
\usepackage{amsmath}

\newcommand{\bfZ}{{\bf z}_{\perp}}

\newcommand{\bfk}{{\bf k}_{\perp}}

\usepackage{xcolor}

\begin{document}
	
\begin{frontmatter}
		
		
		
\title{Gravitational form factors of baryons in a spectator diquark model}

\author{Navpreet Kaur} 
\ead{knavpreet.hep@gmail.com}
		
\affiliation{organization={Department of Physics, Dr. B R Ambedkar National Institute of Technology Jalandhar},
		postcode={144008}, 
		state={Punjab},
		country={India}}
		
\author{Harleen Dahiya} 
\ead{dahiyah@nitj.ac.in}
		
\affiliation{organization={Department of Physics, Dr. B R Ambedkar National Institute of Technology Jalandhar},
		postcode={144008}, 
		state={Punjab},
		country={India}}
		
\begin{abstract}
	Energy momentum tensor (EMT) expresses the interaction between the gravitation and the matter fields, in which the scattering off the graviton is a natural but infeasible probe. However, the EMT can be accessed indirectly through electromagnetic interactions in quantum chromodynamics. The matrix elements of the local EMT operator are parameterized by gravitational form factors, which are subsequently related to the generalized parton distributions. Within the diquark spectator model, we investigate the gravitational form factors of baryons. We consider all the feasible pairs of quark-diquark systems to understand the behavior of each constituent quark flavor of strange and non-strange baryons.
\end{abstract}

		
\end{frontmatter}

\section{Introduction}
	\label{sec1}
	The gravitational coupling of matter is characterized by the energy–momentum tensor (EMT), and the decomposition of the quark EMT in terms of gravitational form factors (GFFs) play a crucial role in understanding the spin structure and spin–orbit correlations of hadrons \cite{Bhoonah:2017olu}. Specifically, the information about the transversely polarized quark can be encoded by studying the current element $T^{\tau \mu \nu}_5$. Burkardt presents the relation between these GFFs and the chiral-odd generalized parton distributions (GPDs), which encode information about the transversely polarized quark structure \cite{Burkardt:2005hp}. The nonforward hadronic matrix elements of lightlike correlation functions of the tensor charge are defined as the chiral-odd GPDs, and there are a total of four chiral-odd GPDs at twist$-$2. Their first Mellin moments at zero momentum transfer refer to the tensor charge and tensor anomalous magnetic moment, and the second Mellin moment leads to the GFFs of transversely polarized quark structure of hadrons.
	
	In the present work, we employed the diquark spectator model  \cite{Bacchetta:2008af} within the light-cone framework to study the GFFs of the proton and a light hyperon $\Xi^0$ to map the difference arising from the different quark structure of these baryons. The spectator diquark model treats a hadron as a two-body system, containing an active quark and a spectator. Based on the diquark spin configuration, the spectator diquark is further characterized as scalar (spin $0$) and axial-vector (spin $1$) diquarks. Several choices of form factors at the baryon-quark-diquark vertex and the polarization states of the diquark are studied in the past years \cite{Gamberg:2007wm, Jakob:1997wg, Bacchetta:2003rz, Brodsky:2000ii}. We have considered a dipolar form for the baryon-quark-diquark vertex to regulate ultraviolet divergences, which is present in the pointlike form. For an axial-vector diquark, summing over the transverse and longitudinal polarization state is considered.
	
\section{Formalism}
For the case of proton and hyperon $\Xi^0$, the instant form wave functions in terms of scalar $\mathfrak{s}$ and axial-vector $\mathfrak{a}$ diquarks can be written as
\begin{eqnarray}
	|p\rangle^{\Uparrow,\Downarrow} &=& \frac{1}{\sqrt{2}} |u ~\mathfrak{s}(ud) \rangle^{\Uparrow,\Downarrow} -\frac{1}{\sqrt{6}} |u ~\mathfrak{a}(ud)\rangle^{\Uparrow,\Downarrow} +\frac{1}{\sqrt{3}} |d  ~\mathfrak{a}(uu)\rangle ^{\Uparrow,\Downarrow},
\end{eqnarray}
\begin{eqnarray}
	| \Xi^{0} \rangle^{\Uparrow,\Downarrow} &=& \frac{1}{\sqrt{2}} |s ~\mathfrak{s}(us)\rangle^{\Uparrow,\Downarrow} +\frac{1}{\sqrt{6}} |s  ~\mathfrak{a}(us)\rangle^{\Uparrow,\Downarrow} -\frac{1}{\sqrt{3}} |u  ~\mathfrak{a}(ss)\rangle ^{\Uparrow,\Downarrow}, \label{Xi} 
\end{eqnarray}
respectively, with $\lambda=(\Uparrow,\Downarrow)$ as their helicity \cite{Lichtenberg:1968zz}. The probabilistic weightage among the scalar isoscalar, vector isoscalar, and vector isovector configurations is 3:1:2. Suppose $p$ and $p^\prime$ are the initial and final state four-vector momentum of baryons and their average is denoted by the term $\bar p$. The light-cone wave functions (LCWFs) \cite{Bacchetta:2008af} for scalar and axial-vector diquarks, with summing over the transverse and longitudinal polarization, can be defined as
\begin{eqnarray}
	\psi^{\lambda}_{\lambda_q} (x,\bfk) = \sqrt{\frac{k^+}{(P-k)^+}} \frac{1}{k^2-m^2_q} \bar{u} (k,\lambda_q) ~\mathcal{Y}_{\mathfrak{s}}~ u(p,\lambda),
	\label{ScalarWfn}
\end{eqnarray}
\begin{eqnarray}
	\psi^{\lambda}_{\lambda_q \lambda_{\mathfrak{a}}}  (x,\bfk) = \sqrt{\frac{k^+}{(P-k)^+}} \frac{1}{k^2-m^2_q} \bar{u} (k,\lambda_q) \epsilon^\ast_\mu (P-k,\lambda_{\mathfrak{a}}) \cdot \mathcal{Y}_{\mathfrak{a}}^\mu~ u(p,\lambda) ,
	\label{VectorWfn}
\end{eqnarray}
where $u(k,\lambda_q)$ corresponds to the spinor for an active quark carrying mass $m_q$, four-vector momentum $k$ and helicity $\lambda_q$. Similarly,  $u(p,\lambda)$ represents the spinor for a baryon with four-vector momentum $p$ and helicity $\lambda$. The polarization of spin $1$ diquark, carrying momentum $(p-k)$ and helicity $\lambda_a$ is represented by the quantity $\epsilon^\ast_\mu (P-k,\lambda_{\mathfrak{a}})$. The scalar $\mathcal{Y}_{\mathfrak{s}}$ and axial-vector $\mathcal{Y}_{\mathfrak{a}}^\mu$ vertices in terms of their respective dipolar form factor $g_\mathfrak{s}$ and $g_\mathfrak{a}$ have the following form 
\begin{eqnarray}
	\mathcal{Y}_{\mathfrak{s}} &=& ig_{\mathfrak{s}}(k^2) \bf{1}, \\
	\mathcal{Y}_{\mathfrak{a}}^\mu &=& ig_{\mathfrak{a}}(k^2) \frac{\gamma^\mu \gamma_5} {\sqrt{2}}.
\end{eqnarray}
The GFFs are defined as the non-forward matrix elements of EMT for tensor current, which can be parameterized by 
\begin{eqnarray}
	\langle p^\prime| \bar{\psi} \sigma^{\tau \mu} \gamma_5 i \overleftrightarrow{D}^\nu \psi |p\rangle &=& \bar{u}(p^\prime,\lambda^\prime) ~ \bigg[ \sigma^{\tau \mu} \gamma_5 \bar{p}^\nu A_{T20}(t) + \frac{\epsilon^{\tau \mu \alpha \beta} \Delta_\alpha \bar{p}_\beta \bar{p}^\nu}{M^2} \bar{A}_{T20}(t) \nonumber \\ &+& \frac{\epsilon^{\tau \mu \alpha \beta} \Delta_\alpha \bar{p}^\nu}{M^2} \gamma_\beta B_{T20}(t) + \frac{\epsilon^{\tau \mu \alpha \beta} \bar{p}_\alpha \Delta^\nu}{M^2} \gamma_\beta  \bar{B}_{T21}(t) \bigg] ~ u(p,\lambda),
\end{eqnarray}
where $A_{T20}(t)$, $\bar{A}_{T20}(t)$, $B_{T20}(t)$ and $\bar{B}_{T21}(t)$ are the GFFs \cite{Burkardt:2005hp, Hagler:2004yt}. The mass of a baryon is represented by $M$. Among $\tau$, $\mu$ and $\nu$ indices, antisymmetrization in $\tau$ and $\mu$ and symmetrization in $\mu$ and $\nu$ is considered. These invariant GFFs can be related to the specific moments of chiral-odd GPDs ($H_T$, $E_T$, $\tilde{H}_T$ and $\tilde{E}_T$) by the following relations 
\begin{eqnarray}
	A_{T20}(t) &=& \int dx~x~H_T(x,\xi,t), \nonumber \\
	\tilde{A}_{T20}(t) &=& \int dx~x~ \tilde{H}_T(x,\xi,t) \nonumber \\
	B_{T20}(t) &=& \int dx~x~E_T(x,\xi,t), \nonumber \\
	-2\xi \tilde{B}_{T21}(t) &=& \int dx~x~ \tilde{E}_T(x,\xi,t). \nonumber
\end{eqnarray}
Here, $x$, $\xi$ and $t$ correspond to the longitudinal momentum fraction, skewness parameter and invariant momentum transfer squared, respectively. Above mentioned chiral-odd GPDs can be evaluated by solving the quark-quark correlator equation with incorporating the tensor current sandwiched between the quark field operators as \cite{Diehl:2005jf}
\begin{eqnarray}
	\frac{1}{2} \int \frac{dz^-}{2\pi} e^{i \bar{x}\bar{p}^+z^-} \bigg\langle p^\prime \bigg| \bar{\psi}\bigg(-\frac{z}{2}\bigg) \sigma^{+j} \gamma_5 \psi \bigg(\frac{z}{2} \bigg) \bigg|p \bigg\rangle \bigg|_{z^+=0, \bfZ=0} &=& \frac{1}{2P^+} \bar{u} (p^\prime,\lambda^\prime) \bigg[H_T(x,\xi,t) \sigma^{+i} \gamma_5 + \tilde{H}_T(x,\xi,t) \frac{\epsilon^{+j \alpha \beta} \Delta_\alpha \bar{p}_{\beta}}{M^2} \nonumber \\&+& E_T(x,\xi,t) \frac{\epsilon^{+j \alpha \beta} \Delta_\alpha \gamma_\beta}{2M} +  \tilde{E}_T(x,\xi,t) \frac{\epsilon^{+j \alpha \beta} \bar{p}_{\alpha} \gamma_\beta}{M} \bigg] u(p,\lambda).
	\label{Corr1}
\end{eqnarray}
The numerical parameters used in the present calculations and the explicit expressions of these chiral-odd GPDs in terms of the overlap form of LCWFs for scalar $\psi^{\lambda}_{\lambda_q}$ and axial-vector $\psi^{\lambda}_{\lambda_q \lambda_{\mathfrak{a}}}$ diquarks are expressed in Ref. \cite{Kaur}. For the case of zero skewness, the term $t$ contains the information about the transverse momentum transfer only. 
		
\section{Results and Discussions}
\begin{figure*}
	\centering
	\begin{minipage}[c]{0.98\textwidth}
		(a)\includegraphics[width=7.0cm]{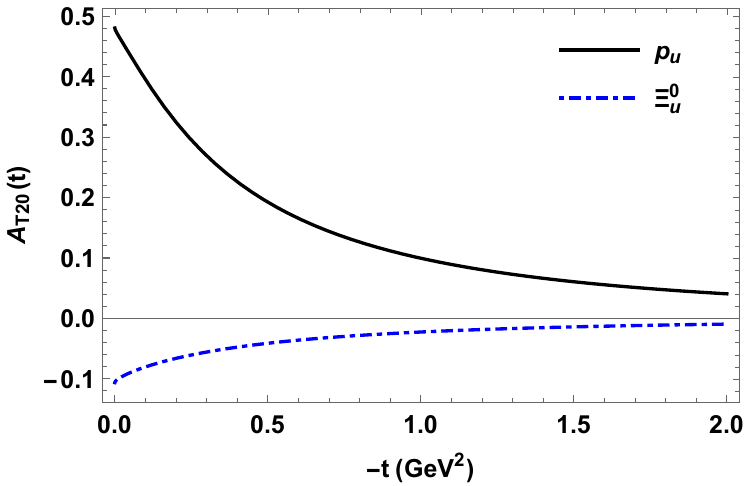}
		\hspace{0.03cm}
		(b)\includegraphics[width=7.0cm]{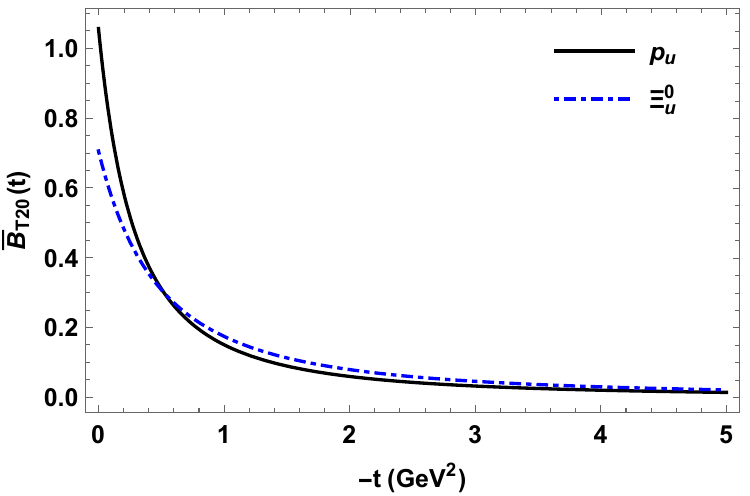}
		\hspace{0.03cm}	 
	\end{minipage}
	\caption{\label{fig1} Comparison of gravitational form factors  (a) $A_{T20}(t)$ and (b) $\bar{B}_{T20}(t)$ as a function of transverse momentum transfer $t$ (GeV$^2$) for $u$ quark flavor of proton and $\Xi^0$.}
\end{figure*}

Fig. \ref{fig1}(a) represents the GFF $A_{T20}$ for the $u$ quark flavor of proton as a function of transverse momentum transfer $t$ (GeV$^2$). At $t=0$, the value of this GFF $A_{T20}$ comes out to be 0.445, which is comparable with the outcome of  basis light-front quantization (BLFQ) calculations, i.e. $A_{T20}(0)=0.480$ \cite{Kaur:2023lun}. Comparison of the contribution of the $u$ quark flavor of proton and light hyperon $\Xi^0$ shows complementary behavior as GFF $A_{T20}$ for $\Xi^0$ has negative values, owing to the flavor dependence of transversity distributions. At $t=0$, the value of GFF is found to be $-0.106$ for the case of $\Xi^0$.

Similarly, comparison of GFF $\bar{B}_{T20}(t)=B_{T20}^{X_q}(t)+2\tilde{A}_{T20}^{X_q}(t)$ for the $u$ quark flavor of proton and $\Xi^0$ is presented in Fig. \ref{fig1}(b). The value of $\bar{B}_{T20}(0)=1.057$, which is little overestimated as compared to the result of BLFQ calculations \cite{Kaur:2023lun}. The $u$ quark flavor of both $p$ and $\Xi^0$ show similar qualitative behavior as a function of $t$ (GeV$^2$). However, the distribution of the $u$ quark flavor $\Xi^0$ is found to decrease slowly as a function of $t$ (GeV$^2$) and has a smaller value at $t=0$ compared to the proton. Succinctly, the $u$ quark flavor in $\Xi^0$ shows a complementary behavior to $p$ for the case of GFF $A_{T20}$. In the forward limit, both GFFs carry smaller values for the $u$ quark flavor of $\Xi^0$ in comparison to the proton.

\section*{Acknowledgments}
	H.D. would like to thank  the Science and Engineering Research Board, Anusandhan-National Research Foundation, Government of India under the scheme SERB-POWER Fellowship (Ref No. SPF/2023/000116) for financial support and International Travel Support (File number: ITS/2025/005034).

\end{document}